\documentclass[twocolumn]{aa}
\usepackage{graphicx}
\def\farcm{\hbox{$.\mkern-4mu^\prime$}}

\begin{document}
\title{Absolute  motions  of  globular  clusters.  II\thanks{Based  on
observations with the NASA/ESA  {\it Hubble Space Telescope}, obtained
at the Space  Telescope Science Institute, which is  operated by AURA,
Inc.,  under NASA contract  NAS 5-26555,  and with  the ESO  {\it Very
Large Telescope} + FLAMES at the Paranal Observatory, Chile, under the
program 71.D-0076}}

\subtitle{{\em HST} astrometry and {\em VLT} radial velocities in NGC 6397}

    \author{Antonino Milone\inst{1},
            Sandro Villanova\inst{1},
            Luigi R.\ Bedin\inst{2},
            Giampaolo Piotto\inst{1},
            Giovanni Carraro\inst{1,3},
            Jay Anderson\inst{4},
            Ivan R.\ King\inst{5},
 	   Simone Zaggia\inst{6}.
           }

    \offprints{L.\ R.\ Bedin}

    \institute{Dip.\ di Astronomia, Univ.\ degli studi di Padova,
          Padova, vic.\ Osservatorio 2, I-35122, EU\\
               \email{milone-villanova-piotto@pd.astro.it}
          \and
              European Southern Observatory, Garching,
          Karl-Schwarzschild-Str.\ 2, D-85748, EU\\
              \email{lbedin@eso.org}
          \and
               Andes Fellow,  Departamento de Astron\'omia, Universidad
 	 de Chile, Casilla 36-d, Santiago, Chile\\
               \email{gcarraro@das.uchile.cl}
 	 \and
          Dept.\  of Physics  and Astronomy,  Mail Stop  108, Rice
          University,  6100  Main  Street,  Houston,  TX  77005,  USA\\
              \email{jay@eeyore.rice.edu}
 	 \and
          Dept.\ of Astronomy, Univ.\ of Washington,
          Box 351580, Seattle, WA 98195-1580, USA\\
              \email{king@astro.washington.edu}
          \and
 	 INAF-Osservatorio Astronomico di Trieste,
 	 via Tiepolo 11, I-34131, EU\\
 	     \email{zaggia@ts.astro.it}
              }

    \date{Received 3 February 2006 / Accepted 27 April 2006}

\abstract{
In this paper we present a new, accurate determination of the three
components of the absolute space velocity of the Galactic globular
cluster NGC 6397 ($l\simeq338^\circ$, $b\simeq-12^\circ$).
We used three {\it  HST}/WFPC2 fields with multi-epoch observations to
obtain astrometric  measurements of objects in  three different fields
in this  cluster.  The identification  of 33 background  galaxies with
sharp nuclei allowed  us to determine an absolute  reference point and
measure  the  absolute  proper  motion  of  the  cluster.   The  third
component has been obtained from radial velocities measured on spectra
from the multi-fiber spectrograph FLAMES at UT2-$VLT$.
We find ($\mu_\alpha  \cos{\delta}$, $\mu_\delta$)$_{\rm J2000.0}$ $=$
($+3.39 \pm 0.15$, $-17.55 \pm  0.15$) mas yr$^{-1}$, and $V_{\rm rad}
= +18.36 \pm 0.09$ ($\pm$0.10) km s$^{-1}$.
Assuming a Galactic potential, we calculate the cluster orbit for
various assumed distances, and briefly discuss the implications.
    \keywords{
astrometry  ---  spectroscopy --- globular clusters:\ NGC 6397 --- dynamics
    } 
    } 
    \titlerunning{Absolute motion of NGC 6397}
    \authorrunning{Milone et al.}
    \maketitle

%
\section{Introduction}
%

After 15 years  of activity, the {\em Hubble  Space Telescope} ($HST$)
is   producing  large   volumes  of   accurate  astrometry   from  the
measurements that the high spatial  resolution of WFPC2 and ACS images
allows.   In   the  archive  are  numerous   fields  with  multi-epoch
observations  that  are  well  sepa\-rated in  time,  and  first-epoch
observations exist  for even more fields.   These are a  gold mine for
proper-motion  measurements  in  a   large  variety  of  Galactic  and
extragalactic  fields,  including  the  crowded  regions  of  globular
clusters (GCs).

Globular clusters are important Galactic tracers; their orbital motion
offers insight into the kinematics and dynamics of the halo, and their
motion provides  important clues  to the distribution  of mass  in the
Galaxy. Moreover,  knowledge of  GC orbits enables  us to  study their
interaction  with the  Galactic disk  and  halo, and  is important  to
dynamical studies of the relationship of the present-day mass function
of a cluster to the initial one.

In the  first paper of  this series (Bedin  et al.\ 2003, Paper  I) we
presented the absolute proper motion of the globular cluster M4, using
as reference  a background QSO.  In  this paper we  use the background
galaxies in the field of another  nearby GC, NGC 6397.  We make use of
archival material,  as well as of  new images taken as  a second epoch
(GO-8656).   We also  use  ground-based spectra  from the  multi-fiber
spectrograph FLAMES  (Pasquini et al.   2002) on UT2-$VLT$  to measure
the  radial   velocity  of  the   cluster.   Finally,  with   the  new
measurements  of proper  motion and  radial velocity,  and  an assumed
Galactic potential, we explore the orbit of the cluster.

%
    \begin{figure}
    \centering
    \includegraphics[width=8.5cm]{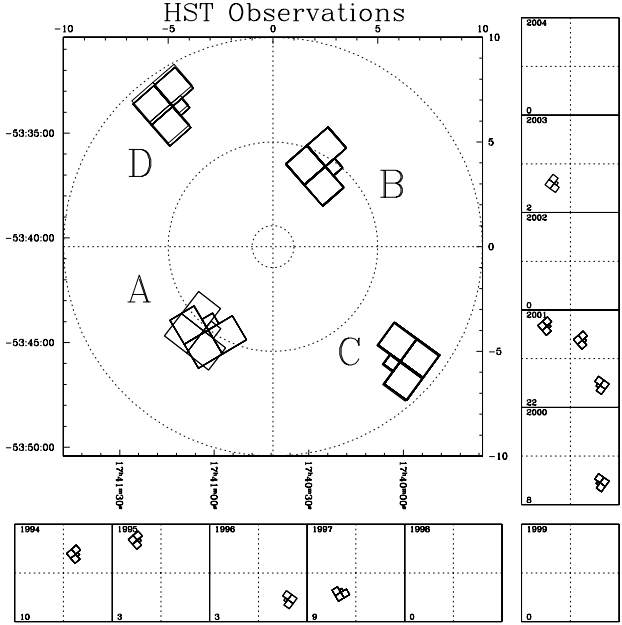}
       \caption{Finding chart  of the WFPC2 fields  considered for this
        work.  The  region is $20'\times20'$  wide, and is  centered on
        the center  of NGC 6397. The  fields measured in  this work are
        labeled `B',`C'  and `D'.  For completeness we  also show field
        `A',  which we  did not  use here  (see text)  but will  use in
        forthcoming  papers.  Fields  A,  B, C,  D  are centered  $5'$,
        4\farcm5,   $8'$,   and   $8'$   from   the   cluster   center,
        respectively.}
          \label{figRAdec}
    \end{figure}
%

%
\section{Observations, data reduction}
%

\subsection{WFPC2/HST:\ Proper motions}

In  Fig.~\ref{figRAdec}  we   show,  in  equatorial  coordinates,  the
footprints of the  three WFPC2 fields of NGC 6397 that  we used in the
present  work. The  center of  the figure  coincides with  the cluster
center.  In the following we will refer  to each field as B, C, and D,
as labeled  in the figure.  Field A  was not used in  the present work
because the two epochs were  oriented differently, and this would have
required a different approach to the reduction, since the chips can no
longer be treated independently.

\begin{table} 
\caption{Description of the data set used in this work.} 
\centering \label{obs} 
\begin{tabular}{ccccc} 
\hline\hline ID & date & EXP-TIME & FILT & GO\\ 
\hline
\multicolumn{5}{c}{ first epoch }\\
\hline

   A  &   Apr'97  & 9$\times$ $\sim$800s                                                                    & F814W & 6797 \\

   B  &   Mar'94  & 3$\times$ $\sim$1000s$+$2100s$+$500s                  & F814W & 5092  \\
      &   Mar'94  & 5$\times$ $\sim$1000s$+$400s                          & F606W & 5092  \\

   C  &   Oct'96  & 3000s$+2\times$1900s                                 & F814W & 6802  \\
      &   Oct'96  & 1300s$+$700s                                         & F814W & 7203  \\
      &   Oct'96  & 1700s$+$400s                                         & F606W & 6802  \\
      &   Oct'96  & 300s                                                 & F606W & 7203  \\

   D  &   Apr'95  & 1800s                                                & F814W & 5369  \\
      &   Apr'95  &         2$\times$1200s                               & F814W & 5370  \\
\hline
\multicolumn{5}{c}{ later epochs }\\
\hline

   A  &    Mar'03 & 500s$+$260s                                          & F606W & 9676 \\

   B  &    Apr'01 & 3$\times$1300s$+$1200s                               & F814W & 8656 \\
      &    Apr'01 & 4$\times$1300s                                       & F606W & 8656 \\

   C  &    Oct'00 & 1200s$+$7$\times$1300s                               & F606W & 8656 \\
      &    Oct'01 & 1200s$+$5$\times$1300s                               & F814W & 8656 \\
   D  &    Apr'01 & 2$\times$1300s                                       & F814W & 8656 \\
\hline 
\end{tabular} 
\end{table} 
Table~\ref{obs} lists for our three fields the exposures that we used.
Almost all  of the image sets  are well dithered, with  both large and
fractional  pixel offsets,  allowing precise  astrometric measurements
and a more accurate assessment of the errors.

%
    \begin{figure}[ht!]
    \centering
    \includegraphics[width=9.0cm]{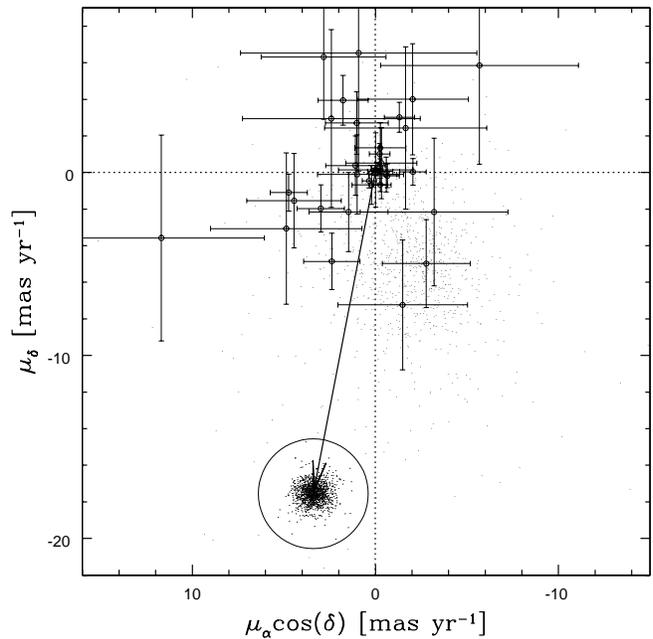}
       \caption{  Vector-point   diagram  of  the   proper  motions  in
       equatorial coordinates, combining  all the fields.  Open circles
       with error bars represent the selected reference galaxies.  Dots
       represent  the stars;  a circle  at 3  mas/yr  separates cluster
       members from  field objects.  The arrow is  the adopted absolute
       proper motion of NGC 6397. }
       \label{vector}
    \end{figure}
%

Positions   were   measured  for   each   filter,   chip,  and   epoch
independently,  by  using  the effective-point-spread-function  (ePSF)
algorithms described by  Anderson \& King (2000).  As  in Paper I, for
proper-motion measurements we used local transformations, and the best
distortion  corrections (Anderson  \& King  2003), after  removing the
34th-row error (Anderson \& King 1999).  The photometry was calibrated
to the  Vega-mag flight  system, following the  recipe of  Holtzman et
al.\ (1995).   For stars with $>$100  DN in their  brightest pixel, we
were able  to measure  positions in  a single image  with an  error of
$<$0.04 pixel in each coordinate.

In  order to  avoid  introducing systematic  errors,  we measured  the
proper motions  from two  epochs in the  same filter. When  two epochs
were available  for each of two  filters (fields B and  C), the proper
motions were obtained from the  weighted mean of the two measurements.
For field B, we linked F814W  (2001) with F814W (1994), and also F606W
(2001) with F606W  (1996); in  field C we  combined F814W  (2001) with
F814W (1996), and  also F606W (2000) with F606W (1996).  In field D we
matched F814W (2001) with F814W (1995).

A visual  inspection of the  images reveals many  background galaxies.
Since 33 of these have  a point-like nucleus, we used our ePSF-fitting
procedure  to measure  their positions.   As was  to be  expected, the
errors found  from multiple measurements of galaxy  positions are much
larger  than  those that  are  typical for  stars,  and  are a  strong
function of the galaxy  morphology.  Nevertheless, those galaxies that
give  good  positions  provide   an  ideal  reference  point  for  the
measurement of absolute proper motions of stars.

Figure~\ref{vector}  shows the  vector  point diagrams  for the  three
fields B, C, and D combined.   Small points show the proper motions of
the stars, and open circles show the motions measured for the selected
galaxies. For  each galaxy we  also show the proper-motion  error bar.
The  axes are parallel  to right  ascension and  declination.  Cluster
members were  used to  compute the transformations  and to  derive the
relative  proper motions  (see Paper  I for  details).  For  the pixel
scale we adopted the number given  by Anderson \& King (2003), and for
the  orientations we  used  information in  the  $HST$ image  headers.
Cluster members constitute the  concentrated distribution of points at
the lower left.  The separation  between field stars and cluster stars
is  very well  defined.  Finally,  the zero  point of  the  figure was
placed at the weighted mean  motion of the galaxies; this converts all
the motions to absolute.
The centroid proper motions of the cluster stars are in good agreement
for the three fields, as is shown in Table~\ref{absPM}.

Taking the  weighted mean of the  positions of the whole  sample of 33
galaxies as zero point of the motion, we find for NGC 6397 an absolute
proper motion, in the J2000.0 system, of
$$\mu_\alpha \cos\delta = +3.39 \pm 0.15\; {\rm mas\; yr^{-1}}, $$
$$\mu_\delta = -17.55 \pm 0.15\; {\rm mas\; yr^{-1}}. $$
In Galactic coordinates this corresponds to 
$$\mu_\ell \cos b = -13.52 \pm 0.15\; {\rm mas\; yr^{-1}},$$
$$\mu_b = -11.70 \pm 0.15\; {\rm mas\; yr^{-1}.}$$ 
The  dominant source of  error is  the uncertainty  in the  zero point
given by the galaxies.

We  note that  the uncertainties  of  the results  for the  individual
fields are quite  different.  This rests in turn on  the wide range of
the measurement  errors for the  individual galaxies, as  indicated by
the lengths  of the error  bars in Fig.~2.   Nearly all of  the weight
comes  from the  few galaxies  that  have the  best measurements;  the
majority  of these  are  in Field  B.   This shows  the importance  of
finding well-exposed galaxies with sharp centers.

These values are in marginal agreement with those found by Cudworth \&
Hanson (1993), which give a proper motion that is $\sim$13\% smaller.
However,  their values  were based  on a  correction from  relative to
absolute proper  motion that was  obtained from a Galactic  model, and
their procedure seems to  provide an analogous discrepancy ($\sim$20\%
smaller) in the case of M4 (Cudworth \& Rees 1990), when compared with
more recent {\em direct}  determinations of its absolute proper motion
(Dinescu, Girard \& van Altena, 1999; and  Paper I).

\begin{table}
\caption{Summary of absolute motion  (in [mas/yr]) for each field. The
          number of background galaxies  used to compute the reference
          point is also indicated. }  
\centering
\label{absPM}
\begin{tabular}{cccccc}
\hline\hline
field   & $\mu_{\alpha} \cos{\delta}$
                  & $\sigma_{ \mu_{\alpha} \cos{\delta} }$
                         & $\mu_{\delta}$
                                 & $\sigma_{\mu_{\delta}}$
                                        & $N_{\rm gal}$  \\
\hline
B       & $+$3.33  & 0.17 & $-$17.53 & 0.17 & 12 \\
C       & $+$3.14  & 0.73 & $-$17.18 & 0.73 &  8 \\
D       & $+$3.75  & 0.40 & $-$17.81 & 0.40 & 13 \\
\hline
average & $+$3.39  & 0.15 & $-$17.55 & 0.15 & 33 \\
\hline 
\hline 
\end{tabular}
\end{table}

\subsection{FLAMES+GIRAFFE/VLT:\ Radial Velocities}

With the aim  of measuring the dispersion of  radial velocities in NGC
6397, in  order to get an estimate  of the distance of  the cluster by
comparing  with the  dispersion of  proper motions  (see Bedin  et al.
2003b for a brief description  of the method), we collected spectra of
1508  stars  at   the  UT2-$VLT$  telescope  in  May   2003  with  the
FLAMES/GIRAFFE spectrograph (proposal No.\ 71.D-0076).  In the present
paper we use these spectra to  measure the mean radial velocity of the
cluster.   Observations  were  performed   in  the  MEDUSA  mode  with
simultaneous  calibration, which  allows obtaining  about  130 spectra
simultaneously  of objects  over a  field of  view of  $25$  arcmin in
diameter.  The  high-resolution set-up H9  (now called H9B)  was used;
this set-up allows a  spectral resolution $R=26,000$ in the 5140--5350
\AA\ range.   This spectral region is well  suited for radial-velocity
measurements because  of the  large number of  metallic lines  and the
absence  of H features.   It also  includes the  Mg triplet,  which is
particularly   important   in   radial-velocity   determinations   for
metal-poor stars  like those of NGC 6397  ([Fe/H] $\sim-$2.0), because
it is the strongest feature well visible also in low $S/N$ spectra.

The target stars for radial  velocities were selected primarily on the
basis of  their position  in the color  magnitude diagram  (CMD).  The
primary criterion used was to  select stars along the red giant branch
(RGB), the subgiant  branch (SGB), and the main  sequence (MS) down to
$V=17.5$,  on a  narrow stripe  $0.08$ magnitude  wide in  color. This
allowed us  to limit drastically  the contamination from  field stars.
Around the turn-off  the stars of NGC 6397 have a  color that is $0.2$
magnitude bluer than  that of the contaminating stars  of the Galactic
disk.  Moreover,  the evolved stars in  NGC 6397 occupy  a position in
the CMD that is well above  the location of the nearby disk stars.  By
analyzing in an outer field, far  from the cluster center, a stripe in
the CMD of  similar size to our selection strip,  we estimated that on
average  only 12 stars  should be  non-members, i.e.,  a contamination
smaller  than 1\%.   (Furthermore, our  radial velocities  showed that
stars in this  part of the CMD are  non-members.)  This left virtually
no contamination in the the sample.

Data  were  automatically   pre-reduced  using  the  GIRAFFE  pipeline
GIRBLDRS\footnote{    Blecha    et     al.     (2000).     See    {\sf
http://\-girbldrs.sourceforge.net/},  for  GIRAFFE pipeline,  software
and  documentation.},  in  which  the  spectra  have  been  de-biased,
flat-field corrected, extracted, and wavelength-calibrated, using both
prior  and  simultaneous   calibration-lamp  spectra.   The  resulting
spectra have  a dispersion  of 0.05 \AA/pixel.   A sky  correction was
applied  to each  stellar  spectrum  by using  the  IRAF {\sf  sarith}
subroutine for subtraction of the average of the sky spectra that were
observed simultaneously (same FLAMES plate) with the stars.

Radial   velocities   were   obtained   from  the   IRAF-{\sf   fxcor}
cross-correlation  subroutine.  Stellar spectra  were cross-correlated
with  a  synthetic template  calculated  by SPECTRUM\footnote{The  LTE
spectral  synthesis program  freely distributed  by Richard  O.\ Gray.
Program       and      documentation      available       at      {\sf
www.phys.appstate.edu/\-spectrum/\-spectrum.html}.},   for   the  mean
temperature, $\log{g}$, and  metallicity appropriate for each observed
star.  The  accuracy of this template  was tested on  a solar spectrum
obtained with the UVES spectrograph; the systematic error is less than
50 m s$^{-1}$.

From multiple observations of 81 stars  it was also possible to make a
direct estimate of the errors.  For stars of magnitude $V=$14, 15, 16,
17,  we  find   errors  of  0.2,  0.3,  0.7,   and  1.4  km  s$^{-1}$,
respectively.

We also searched for possible systematic errors due to misalignment of
the  fibers  or   to  the  difference  in  the   illumination  of  the
simultaneous  calibration   fibers  and   the  fibers  used   for  the
stars\footnote{See   the   GIRAFFE  pipeline   FAQ   web  page,   {\sf
http://girbldrs.sourceforge.net/FAQ/index.cgi}.}.   To   this  end  we
reduced  the daytime wavelength-calibration  plates (the  plates where
all fibers are pointed  at the wavelength-calibration ThAr lamp) using
the  same procedure  that we  used for  our stellar  plates.   Then we
measured  the  velocity shift  of  the  fibers,  using as  a  template
spectrum  the same  ThAr template  that  was used  for the  wavelength
calibration.  As  expected, the average radial  velocity obtained from
the  spectra from  the five  simultaneus calibration  fibers  is zero,
within  the  errors.  By  contrast,  the  radial  velocities from  the
spectra from  the remaining fibers show a  systematic shift, different
for  each fiber,  probably due  to  their incorrect  alignment at  the
entrance of the spectrograph.  For  the purposes of the present paper,
it  is sufficient  to consider  the  mean systematic  shift in  radial
velocity.   We obtained  for the  two positioner  plates Medusa  1 and
Medusa 2:\\

$\Delta V_{\rm rad}\, [{\rm Medusa\ 1, NGC 6397}] = -0.175 \pm 0.009\ {\rm km ~ s^{-1}},$\\

$\Delta V_{\rm rad}\, [{\rm Medusa\ 2, NGC 6397}] = -0.225 \pm 0.008\ {\rm km ~ s^{-1}},$ \\

and the mean value is \\

$\Delta V_{\rm rad} = -0.200 \pm 0.025\ {\rm km ~ s^{-1}},$ \\ 
which must be subtracted from the cluster mean radial velocity.

As a further  test we measured the systematic  error affecting similar
observations  of another  cluster, M4,  which were  collected  a month
later.  In this way we can test the time stability of the spectrograph
by seeing  whether these numbers  remain constant.  We obtained  in an
analogous  way for the  data-set of  M4 the  mean systematic  shift in
radial velocity, for the two  positioner plates Medusa 1 and Medusa 2,
\\

$\Delta V_{\rm rad}\, [{\rm Medusa\ 1, M4}] = -0.176 \pm 0.008\ {\rm km ~ s^{-1}}$ \\

$\Delta V_{\rm rad}\, [{\rm Medusa\ 2, M4}] = -0.236 \pm 0.008\ {\rm
   km ~ s^{-1}}.$ \\

These  values agree  very well  with  the previous  ones.  From  these
results we can assert that GIRAFFE is stable, at least for a period of
about a month.

\begin{figure}[ht!]
    \centering
    \includegraphics[width=9.0cm]{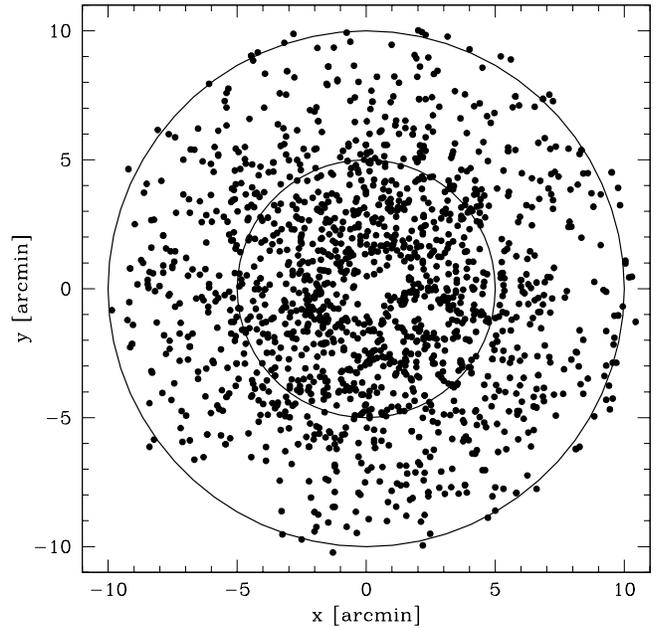}
       \caption{
       Spatial distribution of stars with radial velocity measurement.
       The circles indicate the radial distances of 5 and 10 arc
       min from the center of the cluster.
       }
          \label{vr1}
    \end{figure}

\begin{figure}[ht!]
    \centering
    \includegraphics[width=9.0cm]{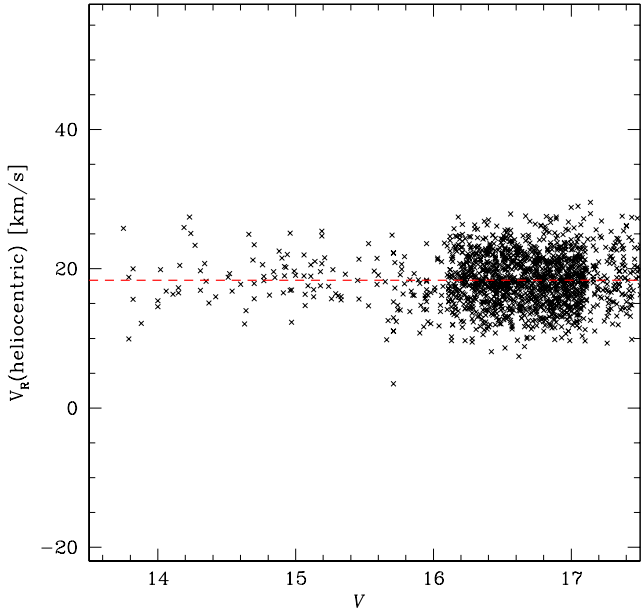}
       \caption{Radial velocities of the observed sample of stars, as
       a function of magnitude.}
          \label{vr2}
    \end{figure}

The spatial  distribution of our stars, and  their heliocentric radial
velocities  as   a  function  of   visual  magnitude,  are   shown  in
Figs.~\ref{vr1} and \ref{vr2}.

%

%
%
%

The unweighted average radial velocity\footnote{
Note that the choice of weights and the calculation of the uncertainty
in the  mean velocity, depend both  on the velocity  dispersion of the
cluster  and on the  uncertainties of  the individual  velocities (see
Pryor \&  Meylan 1993). A much  more careful analysis will  be done in
our  future  work  on  the  internal  dynamics.},  corrected  for  the
systematic shift  previously discussed, and clipped  for outliers more
than 15 km s$^{-1}$ from the mean, is
$$ V_{\rm  rad} = 18.36 \pm 0.09  (\pm 0.03) {\rm km  ~ s^{-1}}.$$ 
The two errors that are given  are first the random error and then the
uncertainty  in the correction  for the  systematic shift.   This mean
comes from 1486 stars (22 were rejected). 

However,  there  are other  effects  of  the  same size  that  afflict
spectroscopic measurements of radial velocities, such as blue-shift of
convective bubbles in giant  stars, or gravitational redshift in dwarf
stars  (Dravins, Lindegren \&  Madsen 1999).   Since we  have observed
both giant and dwarf stars, we add a global systematic error of
0.10 km s$^{-1}$ to our uncertainties. Finally,
$$ V_{\rm  rad} = 18.36 \pm 0.09  (\pm 0.10) {\rm km  ~ s^{-1}}.$$ 
[When  we  analyze the  internal  dynamics  of  the cluster,  we  will
carefully correct for  all these effects, and also  each fiber will be
corrected individually for the calibration bias described above.]

Our  heliocentric  average radial  velocity  for  NGC  6397 is  within
$\sim1.3\sigma$  of the  Da  Costa  et al.\  (1977)  value of  $V_{\rm
rad}=19.8\pm1.1$ km s$^{-1}$, within  $\sim2.1\sigma$ of the Dubath et
al.\ (1997)  value of $V_{\rm rad}=15.1\pm1.6$ km  s$^{-1}$, but there
is    large   disagreement    ($\sim5\sigma$)    with   the    $V_{\rm
rad}=18.80\pm0.10$ km s$^{-1}$ given by Meylan \& Mayor (1991).

Dubath  et al.\ measured  the radial  velocity on  an integrated-light
echelle  spectrum of  the cluster.   Kinematical measurements  on this
kind  of spectra  are known  to  be affected  by strong  observational
biases due to small-number  statistics and strong luminosity weighting
(of  a few red  giants).  As  clearly demonstrated  by Zaggia  et al.\
(1992,  1993)  these biases  influence  both  velocity dispersion  and
radial velocity: systematic  offsets can be as large  as $1\times$ the
velocity dispersion itself.  Therefore it  is not surprising to find a
discrepancy  between  the  integrated-light  radial velocity  and  the
average radial velocity of individual stars.

It is more  problematic to explain the discrepancy  with the Meylan \&
Mayor (1991) radial velocity,  considering that errors in their single
velocities  are similar  to ours,  even  though their  sample is  only
$\sim1/12$  the size  of  ours.   Their measure  is  based on  CORAVEL
spectrometer observations of 127 RGB and SGB stars.  A possible source
of systematic  error could possibly be  in a spectral  mismatch of the
CORAVEL mask  with the low-metallicity  spectra of NGC 6397  stars, as
Meylan  \& Mayor  indeed mention  in  their paper.  Another source  of
systematic error  could be the  presence of unrecognized  field stars,
which, considering  the low  velocity of NGC6397  could bias  the mean
radial velocity.   Meylan \&  Mayor observed mainly  RGB stars  in NGC
6397, while  the majority  of our targets  are turn-off and  MS stars,
which should be relatively free of field contamination.  Unfortunately
a direct  comparison of the radial  velocities of the  two datasets is
not possible.

In  conclusion,  considering  the  work  done to  limit  all  possible
systematics  in the  GIRAFFE spectra,  we  believe that  with a  final
systematic error  $<0.1 $ km s$^{-1}$ and  the large-number statistics
of  nearly 1600  radial  velocities,  our mean  radial  velocity is  a
significant improvement over previous measurements.

\begin{figure}[ht!]
\begin{center}
\includegraphics[width=9.0cm]{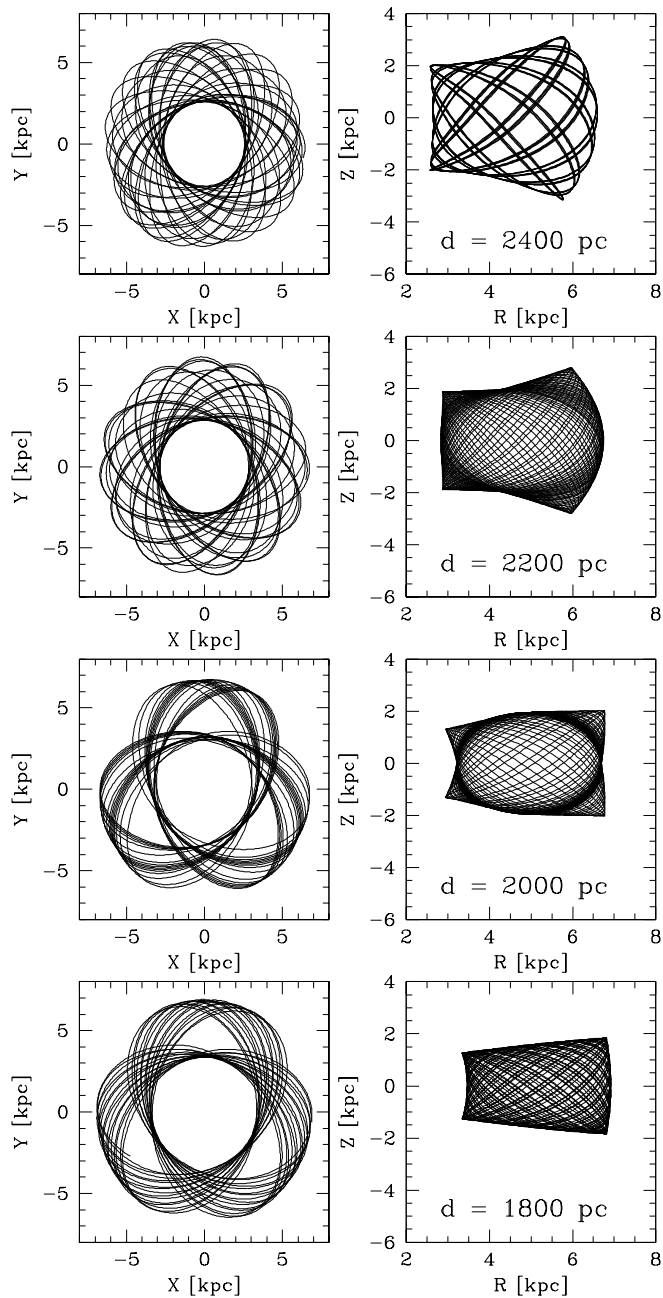}
\protect\caption[]{
Calculated orbits, for assumed distances 1.8, 2.0, 2.2, and 2.4 kpc.
}
\label{orbs}
\end{center}
\end{figure}
%

\begin{table}[!ht]
\caption{Input conditions for orbit calculation. Space velocities
in km s$^{-1}$. } 
\centering
\label{orbIN}
\begin{tabular}{ccccc}
\hline\hline
\multicolumn{5}{c}{ Distance 1.8 kpc }\\
U        & V          & W   $=$ Z  & $\Pi$    & $\Theta$   \\
$34\pm5$ & $-92\pm11$ & $-94\pm10$ & $22\pm5$ & $131\pm11$ \\
\hline
\multicolumn{5}{c}{ Distance 2.0 kpc }\\
U        & V          & W   $=$ Z  & $\Pi$    & $\Theta$   \\
$41\pm5$ & $-103\pm11$ & $-105\pm10$ & $28\pm5$ & $121\pm11$ \\
\hline
\multicolumn{5}{c}{ Distance 2.2 kpc }\\
U        & V          & W   $=$ Z  & $\Pi$    & $\Theta$   \\
$48\pm5$ & $-114\pm11$ & $-116\pm10$ & $35\pm5$ & $111\pm11$ \\
\hline
\multicolumn{5}{c}{ Distance 2.4 kpc }\\
U        & V          & W   $=$ Z  & $\Pi$    & $\Theta$   \\
$55\pm5$ & $-125\pm11$ & $-126\pm10$ & $42\pm5$ & $101\pm11$ \\
\hline
\hline
\end{tabular}
\end{table}

\begin{table}
\caption{Orbit parameters, for the four different distances. 
Units: 
$d${[kpc]}, 
$L_{\rm z}${[kpc km s$^{-1}$]}, 
$E_{\rm tot}${[${\rm 10 \times km^2 s^{-2}}$]}, 
$P${[Myr]}, 
$R_{\rm a}${[kpc]}, 
$R_{\rm p}${[kpc]}, 
$z_{\rm max}${[kpc]}, 
$e${[pure number]}. 
}
\centering
\label{orbOUT}
\begin{tabular}{cccccccc}
\hline\hline
  $d$ & $L_{\rm z}$  & $E_{\rm tot}$& $P$
                     & $R_{\rm a}$& $R_{\rm p}$  &$z_{\rm max}$&  $e$   \\
(1)& (2)& (3)& (4)& (5)& (6)& (7)& (8)\\
\hline
1.8 &  933         & $-$10203       & 112 & 6.92 & 3.34 & 1.84  & 0.35 \\
2.0 &  846         & $-$9899        & 102 & 6.78 & 2.94 & 2.02  & 0.39 \\
2.2 &  766         & $-$9634        &  99 & 6.74 & 2.83 & 2.80  & 0.41 \\
2.4 &  693         & $-$9424        &  93 & 6.59 & 2.57 & 3.11  & 0.44 \\
\hline                                   
\end{tabular}
\end{table}

%
\section{Calculation of the orbital parameters}
%

Our cluster  radial velocity  and absolute proper  motion allow  us to
derive the  three velocity  components of NGC  6397, and  estimate its
Galactic orbit.   The new orbit will  help us, in  principle, to trace
back the dynamical  history of the cluster, clarify  how close it goes
to the Galactic  center, and assess the possible  impact of the motion
on its internal dynamics and its mass function.

Orbit integration for a globular  cluster requires adopting a model of
the potential of  the Milky Way.  We chose that  of Allen \& Santillan
(1991), which was  constructed to fit an assumed  rotation curve and a
distance and velocity of the  Sun with respect to the Galactic Center.
The  potential is  time-independent---clearly  a crude  approximation,
because a significant variation  of the Galactic potential is expected
over  a  typical   globular-cluster  lifetime.   Nevertheless,  it  is
reasonable to believe that the Galactic potential has not changed much
in the last  few Gyr, so that the derived parameters  for the orbit of
NGC 6397, such as the apo- and perigalacticon, can be considered to be
reasonable estimates.

The Galactic  potential of Allen  \& Santillan (1991)  assumes density
distributions in bulge, disk, and halo components.  This gravitational
potential  is  time-independent,  axisymmetric,  fully  analytic,  and
mathematically very  simple.  It has  already been used to  derive the
Galactic orbits of nearby stars  (Allen \& Martos 1986), open clusters
(Carraro \& Chiosi 1994) and disk and halo GCs (Odenkirchen \& Brosche
1992).  The  initial conditions are  given in Table \  \ref{orbIN} for
different cluster distances, and assuming $R_{\circ} = 8$ kpc, and the
value of $V_{\circ} / R_{\circ}$ derived in Paper~I.

The  integration routine  is a  fifteenth-order  symmetric, simplectic
Runge-Kutta,  using the  Radau scheme  (Everhart 1985);  it guarantees
energy  and  momentum conservation  at  the  level  of $10^{-12}$  and
$10^{-9}$, respectively, over the  whole orbit integration. The orbits
have been  integrated back in time for  5 Gyr, and are  shown in Fig.\
\ref{orbs}.

The  orbital parameters  are summarized  in  Table~\ref{orbOUT}, where
column (1) lists the adopted cluster heliocentric distance, column (2)
the  $z$-component  of the  angular  momentum,  column  (3) the  total
energy,  column (4)  the orbital  period  (for radial  motion and  for
vertical motion), columns (5) and  (6) the apo- and peri-center of the
orbit, column  (7) the maximum  vertical distance the  cluster reaches
and  column (8)  the eccentricity,  defined  as $(R_a-R_p)/(R_a+R_p)$.
The  orbit parameters  that we  found do  not differ  much  from those
published by  Dauphole et al.\ (1996)  and by Dinescu  et al.\ (1999),
which means basically  that the Galactic model we  adopt in this paper
is  consistent  with those  used  by  the  previous authors.   Indeed,
Dauphole et al.\ (1996) used an earlier version (Allen \& Martos 1986)
of the Galactic  model we use here (Allen \&  Santillan 1991).  In any
case,  our  newly  derived   space  velocities  make  our  results  an
improvement over previous investigations.   In particular, this is the
first   time   that  proper   motions   calibrated  directly   against
extragalactic  sources have  been used  in studying  the orbit  of NGC
6397.

Independent of  the initial conditions  chosen, the orbit of  NGC 6397
has a  boxy nature.   The cluster oscillates  rapidly through  a dense
part of the  Galactic plane, only a few  kiloparsecs from the Galactic
center. This  kind of  orbit makes NGC  6397 very vulnerable  to tidal
shocks, with a consequent preferential  loss of low mass stars, which,
as first recognized by Piotto, Cool, \& King (1997, see their detailed
discussion),  is  the  likely  cause  of  its  anomalously  flat  mass
function.
 
As the heliocentric distance increases  (from bottom to top panel) the
cluster tends to have a  longer period, to reach greater heights above
the Galactic plane, and to  show a larger epicyclic amplitude, dipping
closer to the Galactic Center.  The assumed distance has a complicated
effect  on tidal  shocks, which  work over  time to  flatten  the mass
function  of the  cluster.  Tidal  shocks  occur three  times in  each
orbital period:\  at perigalactic passage and at  each passage through
the Galactic  plane.  For  a greater assumed  distance the  cluster is
shocked more frequently, and the shocks due to the Galactic center are
stronger,  but the shocks  due to  the plane  occur at  higher cluster
velocity and are thus less effective.

A more  accurate distance determination  is in progress, based  on the
same observing material that is presented  in this paper. It will be a
fundamental step in our understanding of the motion of NGC 6397.

\begin{acknowledgements}
G.\ P.\  and A.\  M.\ acknowledge  the support by  the MIUR  under the
program PRIN2003.   I.\ R.\  K.\ and J.\  A.\ were supported  by STScI
grants GO-8656 and AR-8736. S.\  Z.\ was supported by the MIUR program
PRIN2004. We thank  the referee, Dr C.\ Pryor,  for careful reading of
the manuscript and for many useful comments.
\end{acknowledgements}

%
%


\begin{thebibliography}{}

\bibitem[Allen(1986)]{allen} Allen, C., \&  Martos, M.\ A. 1986, \rmxaa, 13, 137
%
\bibitem[Allen(1991)]{allen91} Allen, C., \&  Santillan, A. 1991, \rmxaa, 22, 255
%
\bibitem[Anderson(1999)]{anderson} Anderson, J., \&  King, I.\ R. 1999, \pasp, 111, 1095
%
\bibitem[Anderson(2000)]{anderson00} Anderson, J., \&  King, I.\ R. 2000, \pasp, 112, 1360
%
\bibitem[Anderson(2003)]{anderson03} Anderson, J., \&  King, I.\ R. 2003, \pasp, 115, 113
%
\bibitem[Bedin(2003)]{bedin}  Bedin, L.\ R., Piotto, G., Anderson, J.,
   \&  King, I.\ R. 2003, \aj, 126, 247, (Paper I)
%
\bibitem[Bedin(2003b)]{Bedi03a} Bedin, L.\ R., Piotto, G., Anderson, J.,
          King, I.\ R. 2003b, ASP Conf.\ Ser.\ 296:\ New Horizons in
          Globular Cluster Astronomy, 296, 360
%
\bibitem[Blecha et al.(2000)]{2000SPIE.4008..467B} Blecha, A., Cayatte, V., North, P., Royer, F., \& Simond, G.\ 2000, \procspie, 4008, 467
%
\bibitem[Carraro(1994)]{carraro} Carraro, G., \& Chiosi, C. 1994, \aap, 288, 751
%
\bibitem[Cudworth(1990)]{cudworth} Cudworth, K.\ M. \& Rees, R. 1990, \aj, 99, 1491
%
\bibitem[Cudworth(1993)]{cudworth93} Cudworth, K.\ M. \& Hanson, R.\ B. 1993, \aj, 105, 168
%
\bibitem[Da Costa et al.(1977)]{1977AJ.....82..810D} Da Costa, G.~S., Freeman, K.~C., Kalnajs, A.~J., Rodgers, A.~W., \& Stapinski, T.~E.\ 1977, \aj, 82, 810
%
\bibitem[Dauphole(1996)]{dauphole} Dauphole, B., Geffert, M., Colin, J., Ducourant, C., Odenkirchen, M. \& Tucholke, H.\ J. 1996, \aap, 313, 119
%
\bibitem[Dinescu(1999)]{dinescu} Dinescu, D.\ I., Girard, T.\ M.  \& van Altena, W.\ F.  1999, \aj, 117, 1792
%
\bibitem[Dravins(1999)]{dravins} Dravins, D., Lindegren, L., Madsen, S. 1999, \aap, 348, 1040
%
\bibitem[Dubath(1997)]{dubath} Dubath, P., Meylan, G., \& Mayor, M.  1997, \aap, 324, 505
%
\bibitem[Everhart(1985)]{everhart} Everhart, E. 1985, 
in Carusi A., Valsecchi G.\ B., eds, Proc. IAU Colloq. 83
Dynamics of Comets: Their Origin and Evolution, Reidel, Dordrecht, p.185 
%
\bibitem[Holtzman(1995)]{holtzman} Holtzman, J.\ A., Burrows, C.\ J., Casertano, S., Hester, J.\ J., Trauger, J.\ T., Watson, A.\ M., \& Worthey, G. 1995, \pasp, 107, 1065 
%
\bibitem[Mayor(1991)]{mayor91} Meylan, G. \&  Mayor, M.  1991, \aap, 250, 113 
%
\bibitem[Odenkirchen(1992)]{odenkirchen} Odenkirchen, M., \& Brosche, P. 1992, Astron. Nachr., 313, 69
%
\bibitem[Pasqu(2002)]{pasq02} Pasquini, L. et al. 2002, Msngr, 110, 1
%
\bibitem[Piotto(1997)]{pio97} Piotto, G., Cool, A.\ M., \& King, I.\ R. 1997 \aj, 113, 1345
%
\bibitem[Pryor(1993)]{pryor93} Pryor, C. \&  Meylan, G. 1993, in Structure and Dynamics of Globular Clusters, ASP, Conference Series, Vol.\ 50, ed.\ S.\ G.\ Djorgovsky \& G.\ Meylan, p.\ 357
%
\bibitem[Zaggia et al.(1992)]{1992A&A...258..302Z} Zaggia, S.~R., Capaccioli, M., Piotto, G., \& Stiavelli, M.\ 1992, \aap, 258, 302
%
\bibitem[Zaggia et al.(1993)]{1993A&A...278..415Z} Zaggia, S.~R., Capaccioli, M., \& Piotto, G.\ 1993, \aap, 278, 415
%
\end{thebibliography}
\end{document}